\documentstyle[amsfonts,epsfig,12pt]{article}
\begin{document}

\title{Breakdown of the Entropy/Area Relationship for NUT-charged Spacetimes}
\author{Dumitru Astefanesei,$^1$\thanks{%
E-mail: {\tt dastefanesei@perimeterinstitute.ca}}~~ Robert B. Mann,$^{2}$%
\thanks{%
E-mail: {\tt mann@avatar.uwaterloo.ca}} ~~Eugen Radu$^3$\thanks{%
E-mail: {\tt radu@thphys.may.ie}} \\
%EndAName
$^{1} ${\small {Department of Physics, McGill University Montr\' eal, Qu\'
ebec H3A 2T8, Canada}}\\
$^{1,2}${\small {Perimeter Institute for Theoretical Physics, Ontario N2J
2W9, Canada}}\\
$^{2}${\small {Department of Physics, University of Waterloo Waterloo,
Ontario N2L 3G1, Canada}}\\
$^{3} ${\small {Department of Mathematical Physics, National University of
Ireland, Maynooth, Ireland}}}
\maketitle

\begin{abstract}
We demonstrate that $\left( D+1\right) $-dimensional, locally asymptotically
anti-deSitter spacetimes with nonzero NUT charge generically do not respect
the usual relationship between area and entropy. \ This result does not
depend on either the existence of closed timelike curves nor on the removal
of Misner string singularities, but instead is a consequence of the first
law of thermodynamics.
\end{abstract}

The relationship between thermodynamic entropy and the area of an event
horizon has simultaneously been one of the more robust and intriguing
concepts in gravitational physics \cite{Beck}. It holds for both black hole
and cosmological event horizons, and has natural extensions to their
dilatonic variants \cite{bharea}. Recovering the entropy/area relationship
is generally regarded as a litmus test that all quantum theories of gravity
must pass \cite{qgrav}.

In a very basic sense gravitational entropy can be regarded as arising from
the Gibbs-Duhem relation applied to the path-integral formulation of quantum
gravity \cite{GibH}, which in the semiclassical limit yields a relationship
between gravitational entropy and other relevant thermodynamic quantities
such as mass and angular momentum. \ In asymptotically de Sitter (dS) or
anti de Sitter (AdS) spacetimes such quantities can be computed
intrinsically, without reference to a background, by employing the
AdS/CFT-inspired counterterm approach \cite{bala}. Whenever it is not
possible to foliate the Euclidean section of a given (stationary) spacetime
by a family of surfaces of constant time, gravitational entropy will emerge %
\cite{Hawking:1998jf}. This situation can occur in $\left( D+1\right) $%
-dimensions if the topology of the Euclidean section is not trivial --
specifically when the (Euclidean) timelike Killing vector $\xi =\partial
/\partial \tau $\ that generates the $U(1)$ isometry group has a fixed point
set of even co-dimension. If this co-dimension is $\left( D-1\right) $ then
the usual relationship between area and entropy holds. \

It has recently become apparent that spacetimes containing a NUT charge
furnish situations in which the co-dimension is smaller than this,
generalizing the relationship between area and entropy \cite%
{Hawking:1998jf,Mann:2003 Found}. Such spacetimes can be constructed by
taking the fibration of a circle over a space of constant curvature. \ If
this curvature is positive (e.g. a product of $p$ $-S^{2}$'s and/or $CP^{2}$%
's) \cite{AwadChamb,Stelea} the orbits of the $U(1)$ isometry group develop
singularities, of dimension $(D-2p)$ in the orbit space, and of dimension $%
(D-2p+1)$ in the Euclidean section. These singularities are the
gravitational analogues of Dirac string singularities and are referred to as
Misner strings \cite{Misner}. If the curvature is non-positive, then Misner
strings are not present \cite{EJM}.

Here we demonstrate that -- regardless of whether or not there are Misner
strings present -- the entropy/area relationship does not hold for
asymptotically AdS spacetimes containing NUT charge. \ This result in stands
in contrast to previous work, in which periodicity $\beta$ of the foliation
for constant positive curvature was adjusted so that singularities from
neither the bolts nor the Misner strings are present, which in turn was
taken to imply that the gravitational entropy is no longer proportional to
the area of the bolts \cite{Hawking:1998jf,EJM,Misentropy}. \ We find
instead that breakdown of entropy/area proportionality is a consequence of
the first law of thermodynamics, which imposes the constraint that the
periodicity of the foliation (interpreted as the inverse temperature) is
proportional to the NUT charge. If the fibration is over a (compact) space
of zero or negative curvature there are no Misner strings, and there are no
constraints on the constant of proportionality. However if the foliation is
over a space of constant positive curvature there will be Misner string
singularities, which can be removed by ensuring that constant of
proportionality is an inverse integer. Our results imply that the
relationship between entropy and area is much less generic than previously
supposed, with Misner strings doing nothing more than quantizing the constant
of proportionality relating NUT charge to inverse temperature.

Since the boundary metric of a $(D+1)$-dimensional Taub-Nut-AdS solution
provides a $D$-dimensional generalization of the known G\"{o}del-type
spacetimes\footnote{These solutions can be embedded in supergravity
theories (of 10 or 11 dimensions ) and may constitute consistent backgrounds
of string theory \cite{susy}.}, our solutions provide a useful tool with
which to investigate the physics in spacetimes containing closed timelike
curves (CTCs) via the AdS/CFT correspondence. Previous investigations have
indicated that such spacetimes are generally believed to be unstable with
respect to quantum fluctuations. However these studies were of spacetimes
with confined causality violation, i.e. with CTCs confined to some region,
with at least one region CTC-free \cite{Visser}. Cauchy horizons separate
the latter regions from the former. This is not the case for the G\"{o}del
spacetime, where causality violation is not a result of the evolution of
certain initial data, but rather has existed ``forever''.

The $(D+1)$-dimensional vacuum Taub-NUT solutions (with $D=2p+1$),
constructed by taking a $U(1)$ fibration over $(M^{2})^{\otimes p}$ have
metrics of the form
\begin{equation}  \label{metric}
ds^{2}=\frac{dr^{2}}{F(r)}+G(r)(d\theta _{i}^{2}+f_{k}^{2}(\theta
_{i})d\varphi _{i}^{2})-F(r)(dt+4nf_{k}^{2}(\frac{\theta _{i}}{2})d\varphi
_{i})^{2},
\end{equation}
with $i$ summed from $1$ to $p$, and $f_{k}(\theta )$ given by%
\[
f_{k}(\theta )=\left\{
\begin{array}{ll}
\sin \theta , & {\rm for}\ \ k=1 \\
\theta , & {\rm for}\ \ k=0 \\
\sinh \theta , & {\rm for}\ \ k=-1,%
\end{array}%
\right.
\]%
where we take the space $M^{2}$ of constant curvature to be positive (a
2-sphere $(k=1)$), zero (a plane $(k=0)$) or negative (a pseudohyperboloid $%
(k=-1)$). By solving the Einstein equation in $(D+1)$ dimensions with
cosmological constant $\Lambda =-D(D-1)/2\ell ^{2}$ , we find $%
G(r)=r^{2}+n^{2}$, while the general form for $F(r)$ is found to be
\[
F(r)=\frac{r}{(r^{2}+n^{2})^{(D-1)/2}}\Big (\int^{r}\left[ k\frac{%
(s^{2}+n^{2})^{(D-1)/2}}{s^{2}}+\frac{D}{\ell ^{2}}\frac{%
(s^{2}+n^{2})^{(D+1)/2}}{s^{2}}\right] ds-2M\Big),
\]%
where the parameter $M$ is an integration constant related to the spacetime
mass. In four dimensions we recover the known form $%
F=(k(r^{2}-n^{2})-2Mr+(r^{4}+6n^{2}r^{2}-3n^{4})/\ell ^{2})/(r^{2}+n^{2})$ %
\cite{Misentropy}.

For generic values of the parameters, the spacetime will have singularities
at $\theta _{i}=\pi $ only in the $k=1$ case. These are the aforementioned
Misner string singularities; they do not appear in the $k=0,-1$ cases. For
any $k$, the Riemann tensor and its derivatives remain finite in all
parallelly propagated orthonormal frames, and no curvature scalars diverge.
The only remaining singularities are (for $k=1$) quasiregular singularities,
which are the end points of incomplete and inextendible geodesics that
spiral infinitely around a topologically closed spatial dimension. World
lines of observers approaching these points come to an end in a finite
proper time \cite{Konkowski}.

To compute the relevant thermodynamic quantities, we analytically continue
in the time coordinate and the NUT charge $t\rightarrow i\tau
,~~n\rightarrow iN$ in order to obtain a solution on the Euclidean section.
For any value of $k$, there will be conical singularities at the roots $%
r_{+} $ of the function $F(r)$ unless the Euclidean time coordinate has the
periodicity%
\begin{equation}
\beta =\left| \frac{4\pi }{F^{\prime }(r_{+})}\right| ,  \label{betarel}
\end{equation}%
which affords a thermodynamic interpretation of $\beta $ as the inverse
temperature \cite{GibH}. Furthermore, the mass parameter $M$ must be
restricted such that the fixed point set of the Killing vector $\partial
_{\tau }$ is a regular at the radial position $r=r_{+}$ (defined by setting $%
F(r_{+})=0$). Two kinds of solutions emerge: ``bolts'', with $r_{+}=r_{b}>N$%
, \ for which the fixed point set is of dimension $(D-1)$,\ and ``nuts'' $%
(r_{+}=N)$, for which the fixed point set is less than this maximal value.
Note that each kind of spacetime has nonzero NUT charge $N$. \

The metrics (\ref{metric}) extremize the gravitational action
\begin{equation}
I_{G}=-\frac{1}{16\pi G}\int_{{\cal M}}d^{D+1}x\sqrt{-g}\left( R-2\Lambda
\right) -\frac{1}{8\pi G}\int_{{\cal \partial M}}d^{D}x\sqrt{-\gamma }\Theta
.  \label{action}
\end{equation}%
The first term in this relation is the bulk action over the $(D+1)$%
-dimensional manifold ${\cal M}$ with metric $g$ and the second term is a
surface term necessary to ensure that the Euler-Lagrange variation is
well-defined, $\gamma $ being the induced metric of the boundary. However,
even at tree-level, the action (\ref{action}) contains divergences that
arise from integrating over the infinite volume of spacetime. These
divergences can be removed by adding additional boundary terms that are
geometric invariants of the induced boundary metric, leading to a finite
total action \cite{bala}. In the context of AdS/CFT \ this action
corresponds to the partition function of the CFT.  An algorithmic
prescription exists \cite{KLS} for computing this additional boundary
action, and the result is
\begin{eqnarray}
I_{{\rm ct}} &=&\frac{1}{8\pi G}\int d^{D}x\sqrt{-\gamma }\left\{ -\frac{D-1%
}{\ell }-\frac{\ell {\sf \Theta }\left( D-3\right) }{2(D-2)}{\sf R}-\frac{%
\ell ^{3}{\sf \Theta }\left( D-5\right) }{2(D-2)^{2}(D-4)}\left( {\sf R}_{ab}%
{\sf R}^{ab}-\frac{D}{4(D-1)}{\sf R}^{2}\right) \right.   \nonumber
\label{Lagrangianct} \\
&&+\frac{\ell ^{5}{\sf \Theta }\left( D-7\right) }{(D-2)^{3}(D-4)(D-6)}%
\left( \frac{3D+2}{4(D-1)}{\sf RR}^{ab}{\sf R}_{ab}-\frac{D(D+2)}{16(D-1)^{2}%
}{\sf R}^{3}\right.   \nonumber \\
&&\left. -2{\sf R}^{ab}{\sf R}^{cd}{\sf R}_{acbd}\left. -\frac{D}{4(D-1)}%
\nabla _{a}{\sf R}\nabla ^{a}{\sf R}+\nabla ^{c}{\sf R}^{ab}\nabla _{c}{\sf R%
}_{ab}\right) +...\right\} ,
\end{eqnarray}%
where ${\sf R}$ and ${\sf R}^{ab}$ are the curvature and the Ricci tensor
associated with the induced metric $\gamma $. The series truncates for any
fixed dimension, with new terms entering at every new odd value of $D$, as
denoted by the step-function (${\sf \Theta }\left( x\right) =1$ provided $%
x\geq 0$, and vanishes otherwise). \ We emphasize that the use and validity
of the counterterm method does not depend upon the validity of AdS/CFT \cite%
{KLS}.

The total action is $I=I_{G}+I_{{\rm ct}}$, and from this one can compute
\begin{equation}
T_{ab}=\frac{2}{\sqrt{-h}}\frac{\delta I}{\delta \gamma ^{ab}},
\label{stressgen}
\end{equation}%
which is a divergence-free boundary stress tensor, whose explicit expression
for $D\leq 8$ is given in ref.~\cite{Das:2000cu}. Provided the boundary
geometry has an isometry generated by a Killing vector $\xi ^{\mu }$, a
conserved charge
\[
{\frak Q}_{\xi }=\oint_{\Sigma }d^{D-1}S^{a}~\xi ^{b}T_{ab},
\]%
can be associated with a closed surface $\Sigma $ (with normal $n^{a}$). If $%
\xi =\partial /\partial t$ then ${\frak Q}$ is the conserved mass/energy $%
{\frak M}$.

Gravitational thermodynamics is then formulated via the Euclidean path
integral
\[
Z=\int D\left[ g\right] D\left[ \Psi \right] e^{-I\left[ g,\Psi \right]
}\simeq e^{-I},
\]%
where one integrates over all metrics and matter fields between some given
initial and final Euclidean hypersurfaces, taking $\tau $ to have some
period $\beta $, determined from eq. (\ref{betarel}) by requiring the
Euclidean section be free of conical singularities. Semiclassically the
total action is evaluated from the classical solution to the field
equations, yielding an expression for the entropy
\begin{equation}
S=\beta ({\frak M}-\mu _{i}{\frak C}_{i})-I,  \label{GibbsDuhem}
\end{equation}%
upon application of the Gibbs-Duhem relation to the partition function \cite%
{Mann:2003 Found} (with chemical potentials ${\frak C}_{i}$ and conserved
charges $\mu _{i}$). The specific heat $C$ is computed from $C=-\beta
\partial S/\partial \beta .$ \ The first law of thermodynamics is then
\begin{equation}
dS=\beta (d{\frak M}-\mu _{i}d{\frak C}_{i}).  \label{1stlaw}
\end{equation}

For $k=1$, an expression for the mass, temperature and finite gravitational
action in arbitrary dimensions was derived in ref.~\cite{Clarkson:2002uj}.
These arguments are straightforwardly extended to any value of $k$, yielding%
\begin{equation}
I_{k,D+1}=\frac{V \beta}{8 \pi G}\left[M -\frac{D}{l^2}\sum_{i=0}^{(D-1)/2}%
\left( {{\frac{(D-1)}{2} }\atop {i }} \right) (-1)^{i}N^{2i}\frac{r_{+}^{(D-2i)}}{%
(D-2i)}\right],
\end{equation}%
where $V=\int \prod_{i=1}^{p}f_{k}(\theta _{i})d\theta _{i}d\varphi
_{i}$ is the total area of the sector $(\theta _{i},\varphi _{i})$
(with $V=(4\pi )^{(D-1)/2}$ for $k=1$). \ The conserved mass-energy
is
\begin{equation}
{\frak M}=\frac{(D-1)VM}{8\pi G},  \label{massgen}
\end{equation}%
where for all nut/bolt solutions the value of $M$ can be computed in terms
of $r_{+}$\ by the condition $F(r=r_{+})=0$
\begin{eqnarray}
M_{k,D+1} &=&\frac{1}{2}\int^{r_{+}}\frac{ds}{s^{2}}\left( (k(s^{2}-N^{2})^{%
\frac{D-1}{2}}+\frac{D}{\ell ^{2}}(s^{2}-N^{2})^{\frac{(D+1)}{2}}\right)
\label{mass-gen} \\
&=&\frac{1}{2}\left( \frac{(r_{+}^{2}-N^{2})^{(D+1)/2}}{l^{2}r_{+}}+(k-\frac{%
N^{2}(D+1)}{l^{2}})\int^{r_{+}}ds\frac{(s^{2}-N^{2})^{(D+1)/2}}{s^{2}}%
\right) .  \nonumber
\end{eqnarray}

Bolt solutions exist for every value of $k$. The condition $F(r_{b}>N)=0$
implies that the $\tau $ coordinate has periodicity
\begin{equation}
\beta =\frac{4\pi r_{b}\ell ^{2}}{k\ell ^{2}+D(r_{b}^{2}-N^{2})},
\label{betabolt}
\end{equation}%
with $r_{+}=r_{b}$ in eq. (\ref{mass-gen}). The action and entropy of the
bolt solutions are given by the general expressions
\begin{eqnarray}
I_{k,D+1}^{b} &=&\frac{V\beta }{16\pi G\ell ^{2}}\left( -\frac{%
(r_{b}^{2}-N^{2})^{(D+1)/2}}{r_{b}}+\bigg(k\ell ^{2}-(D-1)N^{2}\bigg)%
\int^{r_{b}}ds\frac{(s^{2}-N^{2})^{(D-1)/2}}{s^{2}}\right) ,
\label{rel-bolt} \\
S_{k,D+1}^{b} &=&\frac{V\beta }{16\pi G\ell ^{2}}\left( \frac{%
D(r_{b}^{2}-N^{2})^{(D+1)/2}}{r_{b}}+\bigg((D-2)k\ell ^{2}-D(D-1)N^{2}\bigg)%
\int^{r_{b}}ds\frac{(s^{2}-N^{2})^{(D-1)/2}}{s^{2}}\right) ,  \nonumber
\end{eqnarray}%
where the latter quantity is not proportional to the area $%
V(r_{b}^{2}-N^{2})^{(D-1)/2}$ of the bolt.

 Upon insertion of the expressions 
(\ref{mass-gen})-(\ref{rel-bolt}) (viewed as functions of $(r_b,N)$) into the
relation (\ref{1stlaw}),  
we find that for a generic $r_{b}>N$, 
the coefficients of $dr_b$ and $dN$ are nonzero, and so
the first law of thermodynamics fails to be satisfied.
This suggests that the bolt radius is not
independent of the NUT charge and/or the cosmological constant.  We can avoid 
this contradiction by supposing that $r_{b}=r_{b}(N)$ and demanding that these
coefficients vanish.   We then find that the first law of thermodynamics (\ref{1stlaw})
holds if and only if%
\begin{eqnarray}
r_{b\pm }=\frac{\sigma \ell ^{2}\pm \sqrt{\sigma ^{2}\ell
^{4}+D(D+1)^{2}N^{2}\left( DN^{2}-k\ell ^{2}\right) }}{D(D+1)N},
\label{r+bolts}
\end{eqnarray}%
\\
%%%%%%%%%%%%%%%%%%%%%%%%%%%%%%%%%%%%%%%%%%%%%%%%%%%%%%%%%%%%%
\setlength{\unitlength}{1cm}
\begin{picture}(6,6)
\centering \put(-0.3,0){\epsfig{file=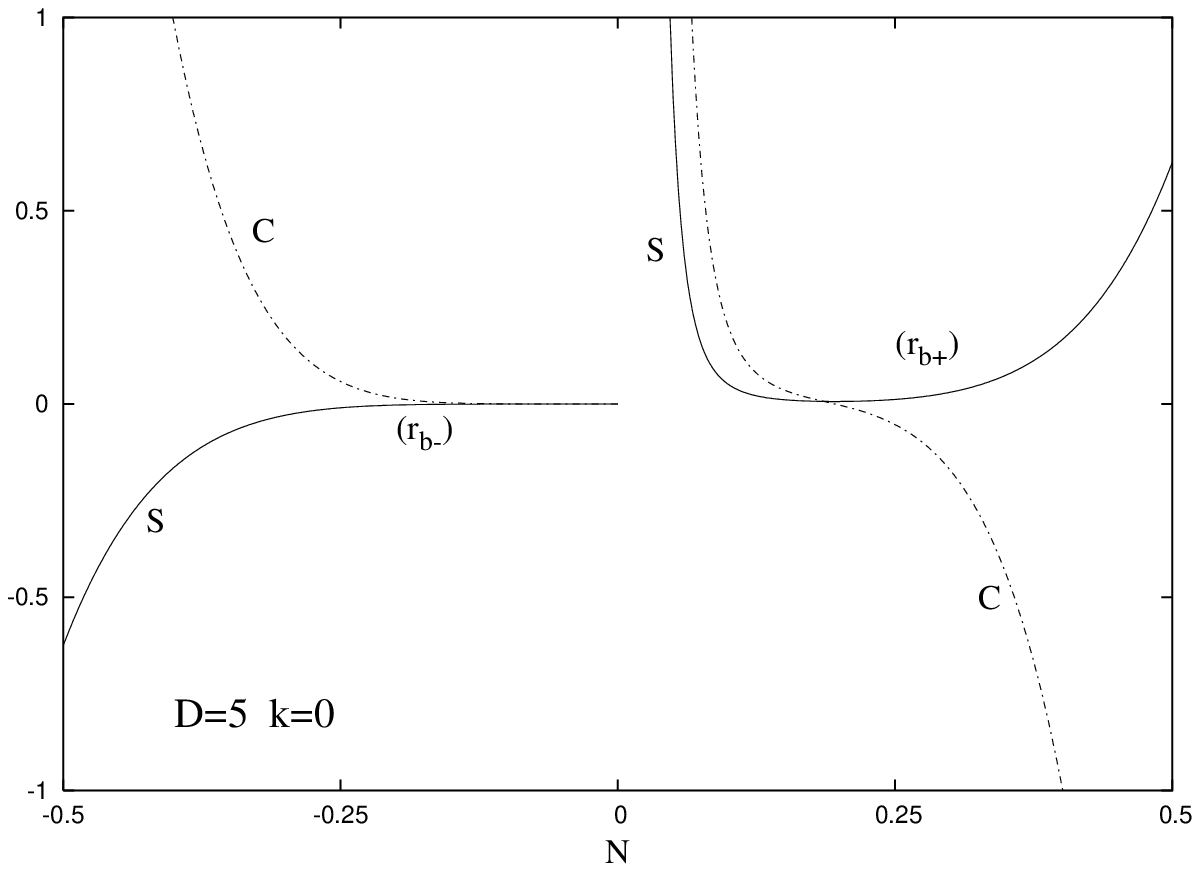,width=9cm}}
\end{picture}
\begin{picture}(-10,1.5)
\centering \put(2.5,0){\epsfig{file=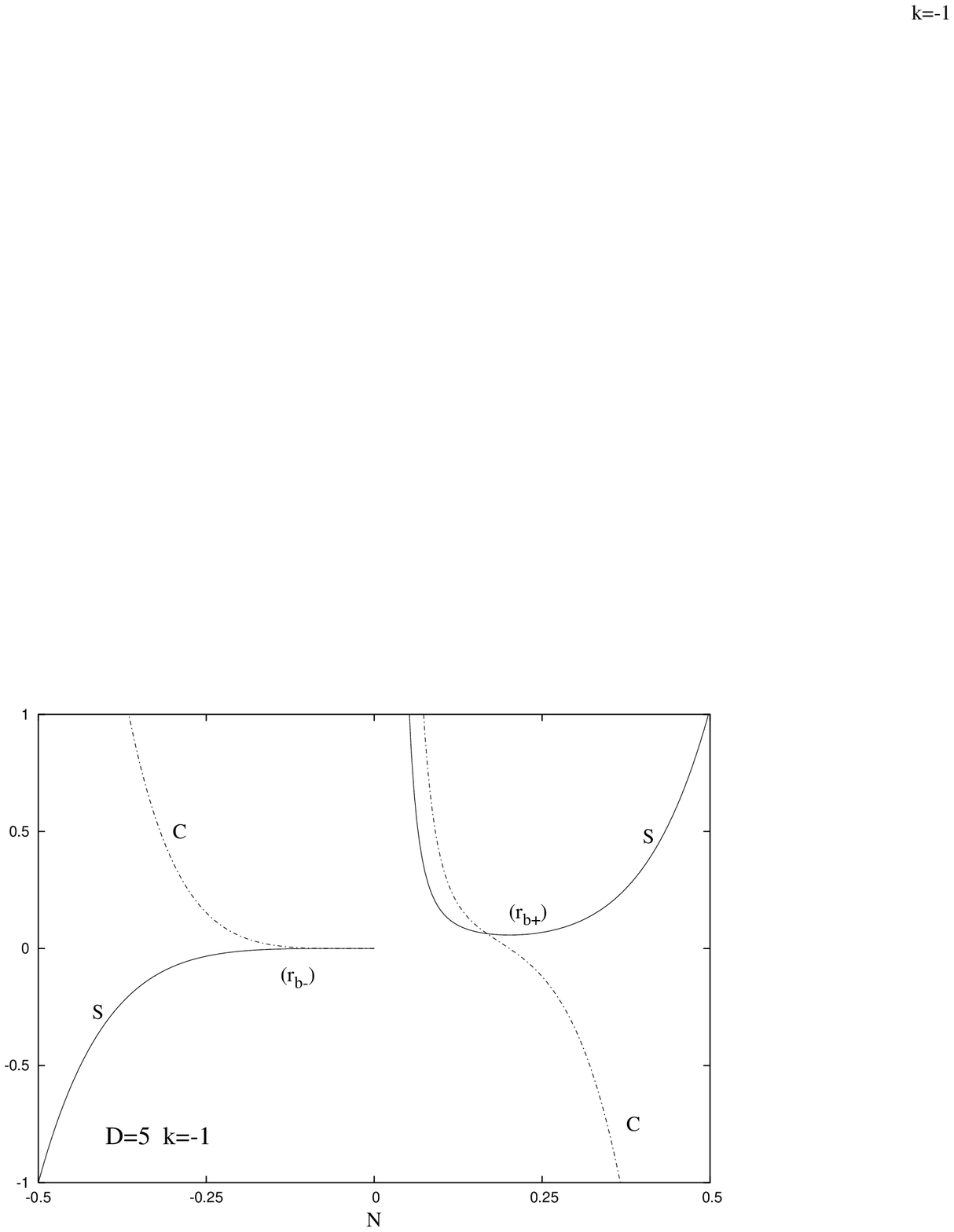,width=9cm}}
\end{picture}
\newline
{\small {\bf Figure 1.} The entropy and specific heat of $k=0,-1$
solutions are represented as a function of $N$ for bolt solutions in
six dimensions. }
\newline
\newline
%%%%%%%%%%%%%%%%%%%%%%%%%%%%%%%%%%%%%%%%%%%%%%%%%%%%%%%%%%%%%%%%%%%%%%%
%
where $\sigma $\ is an arbitrary parameter. 
Insertion of this
relationship into eq. (\ref{betarel}) implies $\beta =2\pi \left(
D+1\right) N/\sigma $. 

 Remarkably, we find that a consistent
thermodynamics forces a relation between the NUT charge and  the bolt
radius that is not imposed either by the field equations or by the regularity
of the geometry.  Indeed, in the  $k=0,-1$ cases
Misner string singularities are absent, and so there are no additional regularity
conditions that can be imposed. However for 
 the $k=1$ spacetimes, regularity demands that these   string singularities 
be removed.  This imposes the additional requirement that 
$\sigma $  be an integer. For all values of $k$, insertion of eq. (%
\ref{r+bolts}) into the expression for the entropy yields an expression that is
not proportional to the area of the bolt unless $N=0$. 

If $k\neq 1$, the parameters $N,\ell $\ may take arbitrary values. For $N>0$%
, we must take $r_{b}=r_{b+}$ (the upper branch) while for negative $N$\ we
must take $r_{b}=r_{b-}$ (the lower branch). \ In this latter case the
condition $r_{b}>N$\ implies that upper limit $|\sigma |<(D+1)/2$. \ If $k=1$
then the situation is more complicated. \ For $\sigma >(D+1)/2$ there are no
constraints on $N$. \ However if $\sigma <(D+1)/2$ then either $N^{2}<\frac{%
\ell ^{2}}{2D}\left[ 1-\sqrt{1-\frac{4\sigma ^{2}}{\left( D+1\right) ^{2}}}%
\right] $ or $N^{2}>\frac{\ell ^{2}}{2D}\left[ 1+\sqrt{1-\frac{4\sigma ^{2}}{%
\left( D+1\right) ^{2}}}\right] $ .

An investigation of the entropy and specific heat of these solutions reveals
a general picture that resembles the $k=1,~D=3$ case. For small $N$, the $%
k=0,-1$ upper branch solutions have positive entropy and specific heat, and
so are thermally stable, whereas the lower branch solutions are always
unstable, with different signs for these quantities. The upper branch
quantities diverge in the high temperature limit while the lower branch
quantities vanish in the same limit. To illustrate this, in Fig. 1, we plot
the entropy and specific heat per unit volume as function of $N$ for the
six-dimensional $k=0,-1$ bolt solutions with $\ell =1,|\sigma|=1$.

It is also possible to consider situations in which $r_{+}=N$. These are the
NUT solutions noted above; the fixed point sets for the class of metrics
that we are considering are $0$-dimensional.\ For $k=-1$, we can prove that
no such solutions exist, generalizing the results found in ref.~\cite%
{Chamblin:1999pz} for $D=3$. \ For $k=0$ we find that for any $D$ the
function $F(r)$ has a double zero at $r_{+}=N$. This implies that the $k=0$
NUT solutions correspond to a background of arbitrary temperature since the
Euclidean time $\tau $ can be identified with arbitrary period.

We can write simple expressions for the parameter $M$, action and entropy of
$k=0,1$ nut solutions
\begin{eqnarray}
M_{k,D+1}^{N} &=&\frac{\sqrt{\pi }}{2\ell ^{2}}(-1)^{(D-1)/2}N^{D-2}\left(
N^{2}(D+1)-kl^{2}\right) \frac{\Gamma (\frac{D+1}{2})}{\Gamma (\frac{D}{2})},
\nonumber  \label{gen1-3} \\
I_{k,D+1}^{N} &=&\frac{V\beta (-1)^{(D-1)/2}}{16\sqrt{\pi }G\ell ^{2}}N^{D-2}%
\Big((D-1)N^{2}-k\ell ^{2}\Big)\frac{\Gamma (\frac{D+1}{2})}{\Gamma (\frac{D%
}{2})}, \\
S_{k,D+1}^{N} &=&\frac{V\beta (-1)^{(D-1)/2}}{16\sqrt{\pi }Gl^{2}}N^{D-2}%
\Big(D(D-1)N^{2}-(D-2)k\ell ^{2}\Big)\frac{\Gamma (\frac{D+1}{2})}{\Gamma (%
\frac{D}{2})},  \nonumber
\end{eqnarray}%
where again $\beta =2\pi \left( D+1\right) N/\sigma $ in order to satisfy
the first law (\ref{1stlaw}). For $k=0$, $\sigma $ is arbitrary and we find
that, for any $D$, there is no region in parameter space for which the
entropy and specific heat are both positive definite.

We close by commenting on our results. We have shown that the breakdown of
the entropy/area relationship arises as a consequence of the first law of
thermodynamics in NUT-charged spacetimes. It does not depend on removal of
Misner string singularities. However the Lorentzian sections of NUT charged
spacetimes typically have closed timelike curves (CTCs) and so one might
expect that our results are a consequence of this feature. \ In fact, this
is not the case. Consider the curve generated by the Killing vector $%
\partial /\partial \varphi $ (for simplicity, we take here $D=3$; the
discussion in the higher dimensional case is similar). We note that%
\[
g_{\varphi \varphi }=4f_{k}^{2}(\frac{\theta }{2})\left(
r^{2}+n^{2}-f_{k}^{2}(\frac{\theta }{2})(4n^{2}F+k(r^{2}+n^{2}))\right) ,
\]%
and so for all $k=1,0$ metrics and those $k=-1$ metrics with $4n^{2}/\ell
^{2}>1$, the curve $r=r_{0},~t=$const$.,~\theta =\theta _{0}\neq 0$ becomes
timelike for large enough values of $(r_{0},~\theta _{0})$, corresponding to
a CTC since $\varphi $ is a periodic coordinate (this type of causality
violation is familiar from the study of the G\"{o}del spacetime \cite%
{Godel:1949ga}).   However there are also regions of these spacetimes that are free
of this type of CTC provided $k\neq 1$ (the $k=1$ metrics necessarily have a periodic time
coordinate $t$ \cite{Misner}).  Specifically, for the $k=-1$ metrics with $%
4n^{2}/\ell ^{2}\leq 1$, the function $h(x^{i})=t$ is a global time
coordinate (i.e. $g^{ij}(\partial h/\partial x^{i})(\partial h/\partial
x^{j})<0$ everywhere). Since these metrics also experience a breakdown of
the entropy/area relationship, the existence of CTCs in the Lorentzian
section cannot be the origin of this phenomenon.

Since in the cases of interest in this paper the dual CFTs are
poorly known (see, e.g., ref.~\cite{Polyakov}), it is impossible to
discuss them in detail. The best one can do at this point is to
compare the (toy example) of scalar field thermodynamics on a
Euclideanized G\"{o}del-type background
with the results obtained on the gravity side \cite{longpaper}. Using eq.~(%
\ref{stressgen}) we obtain in 3 dimensions a $finite$, covariantly conserved
and manifestly traceless dual stress tensor
\[
<\tau _{b}^{a}>=\frac{Mm^{2}}{8\pi G}[3u^{a}u_{b}+\delta _{b}^{a}],
\]%
where $u^{a}=\delta _{t}^{a}$. Similar results hold in higher (odd)
dimensions. As expected \cite{li,stringy1,stringy2}, this situation is
different from the confined causality violating spacetimes, where the energy
momentum tensor of a quantum field diverges at the Cauchy horizon --- unlike
G\"{o}del-type spacetimes, these have geodesic CTCs that dominate quantum
amplitudes in the semiclassical picture. Interestingly enough, using the
zeta function regularization approach a $finite$ Euclidean effective action
for the $k=0$ NUT in $D=3,~5$-dimensions was obtained in ref.~\cite{longpaper}%
. Furthermore, it has the same sign and $similar$ form to the bulk
action (up to a scaling factor $(N/\ell)^{D-1}$) . \ These results
indicate that the boundary CFT is well-defined, and that for this
case the issue of chronology protection cannot be settled at the
level of semiclassical quantum gravity but instead requires a full
quantum gravity theory.

The presence of magnetic-type mass (the NUT parameter $N$) introduces for $%
k=1$ a ``Dirac-string singularity'' in the metric (but no curvature
singularity). We have shown for all $k$ that the presence of a NUT charge
yields additional contributions to the entropy that cause a breakdown of the
entropy/area relationship regardless of the presence/absence of either
Misner strings or CTCs. \ Rather it is a very basic feature of all
NUT-charged spacetimes that any putative quantum theory of gravity will have
to take into account. By extending our solutions to the asymptotically dS
case, we expect that similar results will hold there as well.

\bigskip
%%%%%%%%%%%%%%%%%%%%%%%%%%%%%%%%%%%%%%%%%%%%%%%%%%%%%%%%%%%%%%%%%%%%%%%%%%%
{\Large Acknowledgments} \newline
DA is supported by a Dow Hickson Fellowship. The work of RBM was supported
in part by the Natural Sciences and Engineering Research Council of Canada.
The work of ER was supported by Graduiertenkolleg of the Deutsche
Forschungsgemeinschaft (DFG): Nichtlineare Differentialgleichungen;
Modellierung, Theorie, Numerik, Visualisierung and Enterprise--Ireland Basic
Science Research Project SC/2003/390 of Enterprise-Ireland.

\end{document}